\documentstyle[twocolumn,aps,pre,eqsecnum,psfig,amssymb]{revtex}

\begin{document}

\bibliographystyle{prsty}
\twocolumn[\hsize\textwidth\columnwidth\hsize\csname
@twocolumnfalse\endcsname

\draft

\title{\bf Model for convection in binary liquids}

\author{St.~Hollinger, M.~L\"ucke, and H.~W.~M\"uller}
\address{Institut f\"ur Theoretische Physik, Universit\"at des Saarlandes,
Postfach 151150, D--66041 Saarbr\"ucken, Germany}


\maketitle

\begin{abstract}
A minimal, analytically manageable Galerkin type model
for convection in binary mixtures subject to realistic
boundary conditions is presented. The model 
elucidates and reproduces the typical bifurcation topology of extended 
stationary and oscillatory convective states seen for 
negative Soret coupling: backwards stationary and Hopf bifurcations, saddle 
node bifurcations to stable strongly nonlinear stationary and 
traveling wave (TW) states, and merging of the TW solution branch with 
stationary states. Also unstable standing wave solutions are obtained.
A systematic analysis of the concentration balance for liquid mixture 
parameters has lead to a representation of the concentration field in
terms of two linear and two nonlinear modes. This truncation captures the
important large--scale effects in the laterally averaged concentration field
resulting from advective and diffusive mixing. Also the fact that with 
increasing flow intensity along the TW solution branch the frequency 
decreases monotonically in the same way as the mixing increases --- the
variance of the concentration distribution decreases --- is ensured and 
reproduced well. Universal scaling relations between flow intensity,
frequency, and variance of the concentration distribution (degree of mixing)
in a TW are predicted by the model 
and have been confirmed by numerical solutions of the full equations. The 
validity of the model is checked by comparison with numerical
solutions of the full field equations. 
\end{abstract}


\vskip2pc]
\narrowtext

\newcommand{\Nabla}{\mbox{\bf\boldmath $\nabla$}}
\renewcommand{\vec}[1]{{\bf #1}}
\renewcommand{\u}[1]{\underline{#1}}
\newcommand{\Overrightarrow}[1]{\stackrel{\textstyle\rightarrow}{#1}}


\section{Introduction}
\label{sec:I}
 
A great deal of effort has recently been undertaken to investigate 
convection in binary fluid mixtures as an example for 
pattern formation far from equilibrium \cite{CH93}. This system provides an 
experimentally convenient device 
\cite{WKPS85,OYSK90,TPC96,AR86,LA96,GB86,MS88,EOYSBLKK91,ZM92,WK92,LES96} 
with a well established theoretical description \cite{LL66,PL84} allowing 
quantitative comparisons of theoretical investigations 
\cite{BLHK89,BLK90,BK90,BLKS95I,BPS89,HBL97,KM90,LLT83,LL87,LLMN88,SZ93,V65,Y89,CGLEs,Riecke}
with experiments. For a review and additional references,
see \cite{CH93,BLKS95I}.
Compared to convection in ordinary one--component 
fluids the spatiotemporal properties are far more 
complex due to the influence of Soret sustained concentration 
gradients. The evolution of the concentration field is governed by the 
interplay of typically strong nonlinear convective transport and mixing,
weak dissipative solutal diffusion, and the Soret effect 
\cite{CH93,LL66,PL84}. The latter is a source of concentration 
fluctuations. It generates concentration gradients in response to the 
externally applied temperature difference and to local temperature 
gradients. The strength of the Soret coupling is measured by the 
dimensionless separation ratio $\psi$ \cite{CH93,LL66,PL84}. 

The concentration field 
changes the convective dynamics via solutal buoyancy forces entering 
into the momentum balance. In this way 
concentration gradients directly influence the flow which in turn changes
and mixes the concentration. In binary liquids, this strongly 
nonlinear feed back is only weakly damped by small diffusive 
homogenization so that the concentration 
field distribution shows significant anharmonic
and boundary layer structures. It is, however, the existence of the
feed--back loop that ultimately causes 
convection in binary mixture to exhibit such a rich variety of patterns 
arising from stationary and oscillatory \cite{osc-inst} instabilities:
Depending on the parameters the hydrodynamic balance equations show  
convective solutions that bifurcate out of the quiescent conductive
basic state in the form of (i) straight, stationary, 
parallel rolls, (ii) traveling waves (TWs) consisting of propagating 
rolls, (iii) standing 
wave (SW) oscillations, and (iv) stationary squares. Besides these primary 
states, there are close to onset of convection pulse like, spatially localized 
traveling wave (LTW) states consisting of only few TW rolls, oscillating square
patterns \cite{oscsquares}, cross--roll structures \cite{crossrolls}, 
and also (spiral) defect chaos \cite{spiraldefects}.

In the present paper the focus is on 2--dimensional (2D) spatially extended
convective structures consisting of straight parallel rolls that occur at
negative separation ratios $-0.6<\psi<0$. For  typical fluid parameters
convection arises via an oscillatory subcritical bifurcation. The emerging
solution branch locates unstable TWs that are "weakly nonlinear" only near
the onset. These unstable waves become strongly nonlinear and anharmonic
\cite{HBL97,HL97a} well before the occurrence of a saddle node at which they
are stabilized on an upper solution branch. Simultaneously, the TW propagation
speed slows down from its large value at the Hopf bifurcation threshold
towards zero at the final transition to steady overturning convection (SOC).
There, the amplitude of the concentration wave vanishes since in the SOC state
the fluid is well mixed to a mean  concentration level except within narrow
boundary layers. This SOC state is somewhat similar to  the convective rolls
in one--component fluids.

The bifurcation topology decribed above has been verified by several
experimental groups  (e.~g.~\cite{OYSK90}; for additional references see
\cite{CH93,BLKS95I}). A detailed insight into the spatiotemporal variations
of TW and SOC states along their upper solution branches and their
parameter dependence provided numerical simulations of Barten et al.
\cite{BLHK89,BLKS95I}. A quantitative description of the whole bifurcation
branches \cite{HBL97,HL97a} including the lower branches that were
unavailable to Barten et al. was obtained recently with a
{\sc Multi Mode Galerkin}
(MMG) expansion including several hundred modes. The MMG predictions agree
very well \cite{HBL97,HL97a} with results from finite difference
{\sc Marker And Cell} (MAC) simulations
of the full field equations supplemented by a control process
which allows to evaluate also unstable TW and SOC states \cite{PBprivcomm}. 

Since the first observations of TWs much theoretical research activity has 
been devoted to an understanding and to develop models for these
phenomena. Based upon an earlier few--mode
Galerkin model  \cite{V65} Cross \cite{C86a} and also Ahlers and L\"ucke
\cite{AL87} investigated Soret--driven convection with permeable boundary
conditions. They found TW states only locally at  the onset of convection
since for permeable conditions the onset of TW
convection is tricritical. Linz et al.~\cite{LLMN88} implemented impermeable
conditions yielding a subcritical primary
bifurcation to TWs. This observation is in
agreement with small amplitude computations of Sch\"opf et al.~\cite{SZ93}.
Both approaches however do not explain the stabilization of the TW branch
via a saddle node bifurcation. Bensimon et  al.~\cite{BPS89} considered the
case of weak Soret coupling by means of a small--$\psi$ expansion
treating the concentration field numerically. They
observed a stable TW branch and interpreted the TW--SOC transition as a
boundary layer induced instability. Due to the expansion in
$\psi$ and due to the weak diffusion limit the application range in fluid
parameter space is rather narrow \cite{H96}.

The detailed numerical analyses \cite{BLHK89,BLKS95I,HBL97,HL97a,HL97b}
elucidating the influence of the  spatiotemporal behavior of the concentration
field on various properties of TW states, e. g.,  on the variation of flow
amplitude, frequency, and mixing with heating rate have clearly shown  that
the success of a model description sensitively hinges upon the representation
of the concentration field. It has to
capture the essence of the spatiotemporal structures
following  from the combined action of strong nonlinear advection and weak
diffusion on the one hand  and the generation of Soret induced concentration
currents by temperature gradients on the  other hand. A model that
reproduces with few degrees of freedom all essentials of the  bifurcation
behavior of flow amplitude, frequency, and mixing is presently not
available --- neither in the form of coupled amplitude equations nor in
Galerkin type form. The respective  reasons for their deficiencies are
discussed in the text.

The present paper aims at filling this gap. We present a few--mode
Galerkin model  which rests upon a careful analysis \cite{HL97a,HL97b} of
the concentration balance in liquid mixtures and
explains among others the whole TW solution branch
from oscillatory onset up to its merging  with the upper
SOC branch and the associated changes in the spatiotemporal structure of the
states.

We introduce the system and formulate the theoretical task in
Sec.~\ref{sec:II}. In Sec.~\ref{sec:III} we construct the Galerkin model and
give a detailed account of how the concentration field is represented. The
main body of the paper (Sec.~\ref{sec:IV}) is dedicated  to an extensive
discussion of the results. Wherever possible we provide analytic expressions
for characteristic quantities like thresholds, bifurcation points, and order
parameters like convective  amplitude, frequency, heat flux, and variance of
the concentration distribution. The SOC and TW states will be compared in
quantitative detail with simulations. Our model also yields unstable SW
solutions.


\section{System}
\label{sec:II}

We consider a convection cell of height $d$.
It contains a binary fluid of mean temperature $\bar{T}$ and mean
concentration $\bar{C}$ of the lighter component
confined between two perfectly heat conducting
and impervious plates. This setup is exposed to a vertical gravitational
acceleration $g$ and to a vertical temperature gradient $\Delta T/d$
directed from top to bottom.
The fluid has a density $\rho$ which varies due to temperature and
concentration variations governed by the linear thermal and solutal
expansion coefficients
$ \alpha = - \frac{1}{\rho}\frac{\partial\rho}{\partial \bar{T}} $ and 
$ \beta = - \frac{1}{\rho}\frac{\partial\rho}{\partial \bar{C}} $,
respectively. Its viscosity is $\nu$, the solutal diffusivity
is $D$, and the thermal diffusivity is $\kappa$. The thermodiffusion
coefficient $k_T$ quantifies the Soret coupling which describes the change
of concentration fluctuations due to temperature gradients. 

The vertical thermal diffusion time is used as
the time scale $d^2/\kappa$ of the system and velocities are
scaled by $\kappa/d$. Temperatures are reduced by the temperature
difference $\Delta T$ across the layer and concentration deviations from 
the mean concentration by $\frac{\alpha}{\beta}\Delta T$.
The scale for the pressure is given by $\frac{\rho\kappa^2}{d^2}$.
Then, the balance equations for mass, momentum, heat, and concentration
\cite{LL66,PL84} read in Oberbeck--Boussinesq approximation
\cite{HLL92,BLKS95I}
\begin{mathletters}
\begin{eqnarray}
0 & = & - \Nabla \cdot \vec{u} \label{eq:baleqmass}\\
\partial_t\vec{u} & = & - (\vec{u} \cdot \Nabla) \vec{u}
 - \Nabla\left[p + \left(\frac{d^3}{\kappa^2}g\right) z\right] \nonumber\\
 & & + \sigma \nabla^2\vec{u}
 + R\sigma \left(T+C\right)\vec{e}_z \label{eq:baleqveloc}\\
\partial_tT & = & - \Nabla \cdot \vec{Q}\ =\ 
 - \Nabla \cdot \left[ \vec{u} T - \Nabla T\right]\label{eq:baleqheat}\\
\partial_tC & = & - \Nabla \cdot \vec{J}\ =\
 - \Nabla \cdot \left[ \vec{u} C - L\Nabla\left(C -\psi T\right) \right]\ .
 \label{eq:baleqconc}
\end{eqnarray}
\end{mathletters}
Here, the currents of heat and concentration, $\vec{Q}$ and
$\vec{J}$ respectively, are introduced and $T$ and $C$ denote deviations of
the temperature and concentration fields, respectively, from their global
mean values $\bar{T}$ and $\bar{C}$. The Dufour effect \cite{HLL92,HL95}
that provides a coupling of concentration gradients into
the heat current $\vec{Q}$ and a change of the thermal diffusivity
is discarded in (\ref{eq:baleqheat}) since it is relevant
only in few binary gas mixtures \cite{LA96} and in
liquids near the liquid--vapour critical point \cite{LLT83}.

Besides the Rayleigh number $R=\frac{\alpha g d^3}{\nu \kappa}\Delta T$
measuring the thermal driving of the fluid there enter three additional
numbers into the field equations
(\ref{eq:baleqmass})--(\ref{eq:baleqconc}): the Prandtl number
$\sigma=\nu/\kappa$, the Lewis number $L=D/\kappa$, and the separation
ratio $\psi=-\frac{\beta}{\alpha}\frac{k_T}{\bar{T}}$.
The latter characterizes the sign and the strength of the
Soret effect. Negative Soret coupling $\psi$ induces concentration gradients
antiparallel to temperature gradients. In this situation, the buoyancy
induced by solutal changes in density is opposed to the thermal buoyancy.
When the total bouyancy exceeds a threshold, convection sets in, typically
in the form of straight rolls for negative $\psi$. Ignoring field
variations along the roll axes we describe henceforth 2D convection in an
$x$--$z$ plane perpendicular to the roll axes.

Form and strength of convection and its influence on convective
concentration and temperature transport are measured by the following
order parameters:
(i) The maximum $w_{\rm max}$ of the vertical velocity field.
(ii) The Nusselt number $N = \langle \vec{Q}\cdot\vec{e}_z \rangle_x$
giving the lateral average of the vertical heat current through the system.
In the basic state of quiescent heat conduction its value is $1$ and
larger than $1$ in all convective states.
(iii) The variance
\begin{equation}
\label{eq:Mdef}
M = \sqrt{\langle C^2\rangle_{x,z}/\langle C^2_{\rm cond}\rangle_{x,z}}
\end{equation}
of the concentration field being a measure for the mixing in the system. The
better the fluid is mixed the more the concentration is globally
equilibrated to its mean value $0$ --- optimally mixed, strongly convecting
states enforce $M$ to vanish. In the conductive reference state
denoted by the subscript "cond" the vertical Soret induced concentration
gradient gives rise to a variance of
$\sqrt{\langle C^2_{\rm cond}\rangle_{x,z}} = |\psi|/\sqrt{12}$.
(iv) The frequency $\omega$ of a TW. Thus, extended TWs
with a wave number $k$ have a phase velocity $v=\omega/k$.
They are stationary states in a reference frame co--moving with $v$ relative
to the laboratory system.

The solution of the partial differential equations
(\ref{eq:baleqmass})-(\ref{eq:baleqconc}) requires boundary conditions for
the fields. We use realistic no slip conditions for the top
and bottom plates at $z=\pm 1/2$,
$$ \vec{u}(x,z=\pm 1/2;t) = 0\ \ \ , $$
and assume perfect heat conducting plates by
$$ T(x,z=\pm 1/2;t) = \mp 1/2\ \ \ . $$
Furthermore, impermeability for the concentration is guaranteed by
\begin{equation}
\label{eq:Cbound}
\vec{e}_z\cdot\vec{J} =
 - L\partial_z\left(C-\psi T\right)(x,z=\pm 1/2;t) = 0\ \ \ .
\end{equation}
We should like to stress again that we
restrict ourselves to the description of extended roll
like patterns that are homogeneous in one lateral direction,
say, $y$. So, we investigate 2D states of a certain lateral
periodicity length $\lambda=2\pi/k$. In most cases we take $k=\pi$, i.~e.,
$\lambda$ twice the thickness of the fluid layer,
which is close to the critical wavelengths for the negative
Soret couplings investigated here.

\section{Mode selection and Galerkin model}
\label{sec:III}

\subsection{Temperature and velocity fields}
\label{sec:IIIA}

The temperature field consisting of a linear conductive profile
$-z$ and a convective deviation is truncated by
\begin{eqnarray}
\label{eq:Tansatz}
T(x,z;t) & = -z & + T_{02}(t)\sqrt{2}\sin 2\pi z\nonumber\\
& & + \left[T_{11}(t) e^{-ikx} + \text{c.c.}\right]\sqrt{2}\cos\pi z
\end{eqnarray}
as in the standard Lorenz model \cite{L63} and its first extensions
to convection in binary mixtures with permeable
\cite{V65,C86a,AL87} and impermeable boundaries \cite{LL87,LLMN88,L90}.
These models do not provide a satisfactory representation of
strongly nonlinear TW convection since they used a combination
of concentration and temperature fields in order to fulfill the
impermeability of the plates
exactly without extending the temperature truncation
adequately. For a discussion of this point see Ref.~\cite{HL97a}.
Here, we truncate the concentration itself. This approach avoids the
necessity of a more complicated representation of the temperature field.

For the velocity field we adopt an earlier successful \cite{HL97a}
one--mode description
\begin{equation}
\label{eq:wansatz}
w(x,z,t) = \left[w_{11}(t) e^{-ikx} + \text{c.c.}\right] \cos^2\pi z\ \ \ .
\end{equation}
Eq.~(\ref{eq:wansatz}) completes the Galerkin approximation of velocity
and temperature in our model.

\subsection{Selecting the concentration field modes}
\label{sec:IIIB}

In order to select adequate concentration modes a detailed analysis of the
concentration balance and field structure of SOC and TW states
is necessary.

\subsubsection{Lateral average of the concentration and deviation}
\label{sec:IIIB1}

The first step of our analysis is to decompose the concentration
\begin{equation}
\label{eq:decomp}
C(x,z;t) = \langle C\rangle_x + \Big(C-\langle C\rangle_x\Big)
         =: C_0(z;t) + c(x,z;t)
\end{equation}
into its lateral mean $C_0(z;t)$ and the deviation
$c(x,z;t)$ from it.
Inserting this decomposition into the balance equation
(\ref{eq:baleqconc}) for the concentration
and averaging it yields two coupled equations for $C_0$ and $c$
\begin{mathletters}
\begin{eqnarray}
\partial_t C_0 
 & = & -\partial_z\langle w c\rangle_x + L\partial_z^2 C_0
  \label{eq:baleqC0}\\
\left(\partial_t + \vec{u}\cdot\Nabla \right) c 
 & = &     \partial_z\langle w c\rangle_x
       - w \partial_z C_0 + L\nabla^2 c\ \ . \label{eq:baleqkleinc}
\end{eqnarray}
\end{mathletters}
In both of these equations we have discarded the Soret coupling term
$L \psi \Nabla^2 T$ {\em in the bulk\/} of the fluid. However,
the Soret coupling will not be dropped in the boundary condition
(\ref{eq:Cbound}). The motivation and justification
for this approximation is discussed in quantitative detail in
Ref.~\cite{HL97a}. Here, we only mention that
the basic justification is the smallness of the
Lewis number $L$ in liquids so that transport by diffusion an the Soret 
effect --- both
enter the balance with a weight $L$ --- are small compared with advection.
In those regions where advection needs to be balanced by another transport
mechanism strong concentration gradients are observed whereas the
temperature gradient shows no such boundary layers. Thus, an adequate
balance is assured by advection and diffusion and the additional
concentration source or sink --- the Soret effect ---
can be omitted in the bulk. Only
in the impermeable boundary condition the Soret effect leads to a sizeable
nonvanishing mean concentration gradient at the plates which cannot be
ignored. 

In a SOC fixed point and also in a TW the lateral average
of the concentration field is temporally constant. Thus, it can be
calculated explicitly from (\ref{eq:baleqC0}) to be
\begin{equation}
\label{eq:C0erg}
C_0(z) = -\psi N z +
\frac{1}{L}\int_0^z\!dz'\,\langle w c\rangle_x\ \ .
\end{equation}
Here, the impermeable boundary condition (\ref{eq:Cbound})
\begin{equation}
\partial_zC_0(\pm1/2) =
\psi\,\partial_zT_0(\pm1/2)\ = \ -\psi\,N\label{eq:zeroboundconc}
\end{equation}
relating the lateral averages $C_0$ and $T_0$ has been used in the first
integration of (\ref{eq:baleqC0}) from $-1/2$ to $z$. The second
integration is taken from $0$ to $z$ since $C_0(z=0)=0$ as
required by the mirror glide symmetry $C(x+\lambda/2,z)=-C(x,-z)$ for SOC
and TW states \cite{BLHK89,BLKS95I}. When
describing TW as well as SOC fixed points the relation (\ref{eq:C0erg})
can be inserted into the evolution equation (\ref{eq:baleqkleinc}) giving
\begin{equation}
\label{eq:cerg}
(\partial_t + \vec{u}\cdot\Nabla)\,c 
 =   L\,\nabla^2\,c + \psi N w
   + \left(\partial_z - \frac{w}{L}\right)\langle w c\rangle_x\ \ .
\end{equation}
The solution $c$ of this equation completely determines the relaxed TW and
SOC concentration
field whenever the vertical velocity field together with the Nusselt number
is given. Just that is realized by the truncations for the velocity
(\ref{eq:wansatz}) and temperature (\ref{eq:Tansatz}) field in
Sec.~\ref{sec:IIIA}. With $c$, $w$, and $N$ also the TW and SOC
spatial structure of
the lateral mean of the concentration is determined via (\ref{eq:C0erg}).
The last task is therefore to select modes for $c$ that approximate the TW
and SOC structure appropriately.

\subsubsection{Symmetry decomposition and lateral mode truncation}
\label{sec:IIIB2}

The mode selection of $c$ is based among others on the insight gained in
Ref.~\cite{HL97b} from a symmetry decomposition of $c$. 
The decomposition was realized \cite{HL97b} with respect to
the different parities under the mirror operations $x\rightarrow -x$ and
$z\rightarrow -z$ so that the four symmetry classes
${\cal S}^{++}$, ${\cal S}^{+-}$, ${\cal S}^{-+}$, and ${\cal S}^{--}$ are
obtained. The first (second) superscript denotes the parity under the
operation $x\rightarrow -x$ ($z\rightarrow -z$). In relaxed TWs and
SOCs the lateral coordinate $x$ can be combined with the time $t$
to $x-vt$ so that the time derivatives in TWs can be replaced by
$-v\partial_x$ and the argument $t$ in the field $c$ can be dropped.
One main result of Ref.~\cite{HL97b} was the observation that field
components of the symmetry ${\cal S}^{--}$ are not needed for a
quantitative description of the TW and SOC
bifurcation topology and that the fields
of the class ${\cal S}^{+-}$ are made up mainly by the zeroth lateral Fourier
mode. Consequently, the  SOC and TW concentration field $c$ can be
represented well by just two parts
\begin{equation}
\label{eq:cppmprestrict}
c(x,z) = c^{++}(x,z) + c^{-+}(x,z)
\end{equation}
belonging to the symmetry classes ${\cal S}^{++}$ and
${\cal S}^{-+}$. Using the approximation (\ref{eq:cppmprestrict})
for $c$ one obtains from (\ref{eq:cerg}) two equations
\begin{mathletters}
\label{eq:cppmpgleich}
\begin{eqnarray}
-v\,\partial_x c^{-+} + \frac{1}{L}w\,\langle w c^{++}\rangle_x
 & = & N\psi\,w\ + L\,\nabla^2 c^{++}\\
-v\,\partial_x c^{++} 
 \phantom{+ \frac{1}{L}w\,\langle w c^{++}\rangle_x}
 & = & \phantom{N\psi\,w\ + } L\,\nabla^2 c^{-+}
\end{eqnarray}
\end{mathletters}
for the $c^{++}$ and $c^{-+}$ fields of fully relaxed SOCs and TWs. Here,
the vertical velocity field $w$ was fixed to belong to
the symmetry class ${\cal S}^{++}$ in the ansatz
(\ref{eq:wansatz}) by choosing the temporal phase adequately and then
switching from $x$ to $x-vt$. Furthermore, we used the fact that the
application of the advective derivatives $u\partial_x$ and $w\partial_z$ to
$c^{++}$ and $c^{-+}$ generate fields with negative vertical parity that do
not belong to the two retained symmetry classes ${\cal S}^{++}$ and
${\cal S}^{-+}$ for $c$. The same holds also for
$\partial_z \langle w c \rangle_x = \partial_z \langle w c^{++} \rangle_x$.

The important implication
of Eqs.~(\ref{eq:cppmpgleich}) is that the 
{\em lateral variation\/} of the concentration
field $c(x,z)$ is restricted to $\sin\,kx$ and $\cos\,kx$ if one uses the
approximations (\ref{eq:cppmprestrict}) and (\ref{eq:wansatz}). The reason
is that $w\,\langle w c^{++}\rangle_x$ as well as $N\psi\,w$ have
the lateral variation of $w$, i.~e., $\cos\,kx$, and no other inhomogeneity
in (\ref{eq:cppmpgleich}) excites higher modes. 
We should like to stress that all these restrictions are
based on a quantitative investigation of their implications. Thus, they
do not endanger the success of our model as we will see below.

\subsubsection{Vertical variation}
\label{sec:IIIB3}

The last task is to select modes for the
{\em vertical spatial dependence\/} of $c(x,z)$ in the form
\begin{equation}
c(x,z) = c_1(z)e^{-ikx}+\text{c.c.}
\end{equation}
with $c_1$ being the first lateral Fourier mode of $c$. Numerical
calculations \cite{BLKS95I,HL97a} have revealed that the main contribution
to $c_1(z)$ is made up by a part being phase shifted by $90^{\rm o}$ with
respect to the vertical velocity field. With the velocity
ansatz (\ref{eq:wansatz}) and the choice that $w_{11}$ is real,
this implies that $c_1(z)$ is dominated by
its imaginary part. This holds as long as the phase velocity $v$ is
large compared with the Lewis number $L=O(0.01)$.

\begin{figure}[t]
\centerline{\psfig{figure=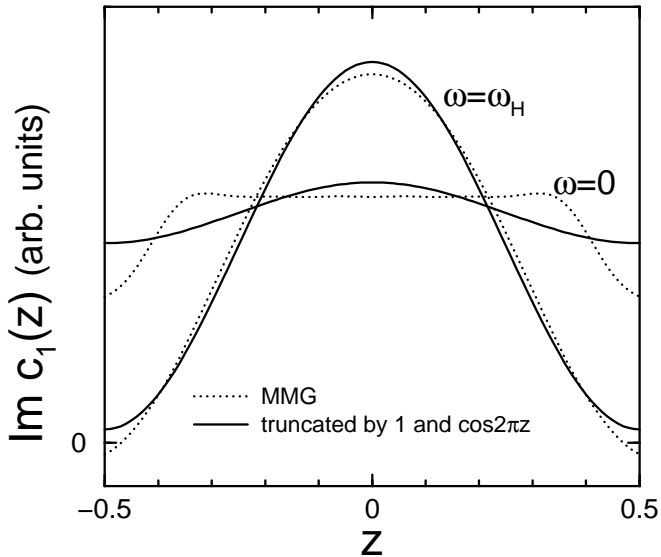,width=\linewidth}\vspace*{3mm}}
\caption{Vertical profiles of that part of the first lateral Fourier mode of
         the concentration being in phase with the streamfunction (in our
         notation the imaginary part). Shown are the eigenfunctions
         at the Hopf bifurcation ($\omega=\omega_{\rm H}$) and at
         the SOC--TW transition ($\omega = 0$) with arbitrary
         normalizations. Dots refer to a many mode
         Galerkin scheme, solid lines to an optimal truncation
         by a linear combination of $1$ and $\cos\,2\pi z$.
         Parameters are $L=0.01$, $\sigma=10$, and $\psi=-0.25$.}
\label{fig:trunc}
\end{figure}
The vertical spatial
dependence of $\mbox{Im}\,c_1(z)$ is investigated in Fig.~\ref{fig:trunc}.
Therein, we demonstrate that the "exact" solution (dotted line) obtained from
a MMG scheme can very well be reproduced by a
linear combination of $1$ and $\cos\,2\pi z$, i.~e., of the first two modes
with the boundary condition $\partial_zc = 0$ at $z=\pm1/2$. This holds
both for the bifurcation out of the heat conducting state
($\omega=\omega_{\rm H}$) and for the SOC--TW transition ($\omega=0$).
The mode amplitudes of $1$ and $\cos\,2\pi z$ were optimally chosen
in Fig.~\ref{fig:trunc} in order to
demonstrate the usefulness of the representation of the first lateral
Fourier mode
\begin{equation}
c_1(z;t) = \sqrt{2}\,\left[c_{10}(t) + c_{12}(t)\cos\,2\pi z\right]
\label{eq:c1ansatz}
\end{equation}
in terms of $1$ and $\cos\,2\pi z$ with two
complex amplitudes $c_{10}(t)$ and $c_{12}(t)$.

According to
Eq.~(\ref{eq:C0erg}), the {\em spatial dependence\/} of the zeroth Fourier
mode $C_0(z)$ can be calculated by integrating the product of $\cos^2\pi z$
for the vertical dependence of $w$ with $1$ and $\cos\,2\pi z$,
respectively, coming from $c$. This procedure leads to vertical modes
of the form
$(2\pi z + \sin\,2\pi z)$ and $(-4\pi z + \sin\,4\pi z)$. Their real
amplitudes are labelled by $c_{02}$ and $c_{04}$, respectively. These two
modes $c_{02}$ and $c_{04}$ are constant in SOC and TW fixed points.
To derive a model that includes temporal variations of these modes as well,
both modes are taken as time dependent.

The complete Galerkin ansatz 
for the concentration field is therefore given by
\begin{eqnarray}
\frac{C(x,z,t)}{-\psi} & = & \phantom{ + }
\left\{1+2\pi\sqrt{2}\big[c_{02}(t)-2 c_{04}(t)\big]\right\}z\nonumber\\
& & +\ c_{02}(t) \sqrt{2}\sin 2\pi z + c_{04}(t)\sqrt{2}\sin 4\pi z\nonumber\\
& & +\ \left[c_{10}(t) e^{-ikx} + \text{c.c.}\right]\sqrt{2}\nonumber\\
& & +\ \left[c_{12}(t) e^{-ikx} + \text{c.c.}\right]\sqrt{2}\cos 2\pi z\ \ .
\label{eq:Cansatz}
\end{eqnarray}
In order to avoid the appearence of
temperature modes in (\ref{eq:Cansatz}) we approximate the
boundary condition (\ref{eq:zeroboundconc}) by
$\partial_zC_0(\pm 1/2)=-\psi$ that deviates from the correct value by a
factor equal to the Nusselt number $N=O(1)$.
This approximation can be understood as the leading term in an amplitude
expansion of $N$ which starts at $N=1$. The exact value of $N$ is of minor
importance in the boundary condition (\ref{eq:zeroboundconc}). Only
the existence of a finite slope of $C_0$ at the plates is crucial.

Similarly, a lateral variation of the
vertical derivative of $C$, i.~e., of $c$, at the plates can be seen as a
higher order contribution that scales with the field amplitudes and not with
$O(1)$. The reader can convince himself of the smallness of the derivatives of
$\mbox{Im}c_1(z)$ at the plates in Fig.~\ref{fig:trunc}.

\subsection{Galerkin model}
\label{sec:IIIC}

\subsubsection{Scalings}
\label{sec:IIIC1}

We use $r = \frac{R}{R^0}$ as control parameter. Here,
$R^0 = \frac{1}{6}\left(\frac{3\pi}{2}\right)^6 \simeq 1825.14$ is the
stability threshold of the quiescent heat conducting state of the pure
fluid with respect to disturbances of a wave number $k=\pi$ within our
model. This is not exactly the minimum of the marginal curve. It is
calculated as $0.9998 R^0$ at $k = 0.9827 \pi$. But since we are not
interested here in wave number depedencies we fix $k=\pi$.

The complex amplitudes of the first lateral Fourier modes, 
$f\in\left\{w_{11},T_{11},c_{10},c_{12}\right\}$, are written in a
vector notation
\begin{equation}
{\rm\bf f} =
\left(\mbox{Re}\,f\ ,\ \mbox{Im}\,f \right)^T.
\end{equation}
We scale the mode
amplitudes in the following way:
\begin{mathletters}
\begin{eqnarray} 
\label{eq:scalex}
{\bf X} & = & \frac{8}{5\pi^2}{\rm\bf w_{11}}\ \ ,\\
\label{eq:scaleyz}
{\bf Y} = \frac{6\pi\sqrt{2}}{5}\,r{\rm\bf T_{11}}
 & \ \  ,\ \ & Z = \frac{6\pi\sqrt{2}}{5}\,r\,T_{02}\ \ ,\\
\label{eq:scaleu1u2}
{\bf U_1} = \frac{32\sqrt{2}}{5}\,r\,{\rm\bf c_{10}}\ \ & , &
{\bf U_2} = \frac{32\sqrt{2}}{5}\,r\,{\rm\bf c_{12}}\ \ ,\\
\label{eq:scalev1v2}
V_1 = \frac{256\sqrt{2}}{15\pi}\,r\,c_{02}\ \ & , &
V_2 = \frac{256\sqrt{2}}{5\pi}\,r\,c_{04}\ \ .
\end{eqnarray} 
Additionally, we introduce
\begin{equation}
\label{eq:taudef}
\tilde{\sigma}=\frac{27}{14}\sigma\ \ ,\ \
   \tau=\frac{1}{2\pi^2}\ \ ,\ \
   a = \frac{9\pi^2}{128}\simeq 0.6940\ \ \ .
\end{equation}
\end{mathletters}

\subsubsection{Order parameters}
\label{sec:IIIC2}

The order parameters maximal convective amplitude $w_{\rm max}$, Nusselt
number $N$, and mixing number $M$ can be expressed by
\begin{mathletters}
\begin{eqnarray}
w_{\rm max}
  & = & \frac{5\pi^2}{4}\mid{\bf X}\mid\ \ ,\\
N & = & \int_{-1/2}^{1/2}\!\!\! dz \langle\vec{Q}\cdot\vec{e}_z\rangle_x\ =\ 
        1+\frac{25}{18r}{\bf X}\!\cdot\!{\bf Y}\ \ ,\\
M^2 & = & \frac{12}{\psi^2}\,<C^2>_{x,z}\nonumber \\
    & = & 1 + \frac{75}{128\,r^2}
     \left(|{\bf U_1}|^2+\frac{1}{2}|{\bf U_2}|^2\right)\nonumber\\
    & & \phantom{1} + \frac{45}{64\,r}\left(1+\frac{\pi^2}{3}\right) V_1\
        -\ \frac{15}{128\,r}\left(1+\frac{4\,\pi^2}{3}\right) V_2\nonumber\\
    & & \phantom{1} + \frac{1125\,\pi^2}{32768\,r^2}\left(V_1-\frac{1}{3}V_2\right)^2\nonumber\\ 
    & & \phantom{1} + \frac{75\,\pi^4}{16384\,r^2}\left(V_1-\frac{2}{3}V_2\right)^2\ \ .
\end{eqnarray}
\end{mathletters}
Here, the Nusselt number is computed as the global
spatial average of the vertical heat flux since due to
the truncation of velocity and temperature fields in different bases the
laterally averaged vertical heat flux which is conventionally
used for evaluating the
Nusselt number has a slight $z$--dependence. This problem occurs in
all few mode Galerkin approximations with no slip boundary conditions, see,
e.~g., \cite{NLK91,HL95}.

\subsubsection{Model}
\label{sec:IIIC3}

We insert the field truncations of Secs.~\ref{sec:IIIA} and
\ref{sec:IIIB} into the basic equations
(\ref{eq:baleqveloc})--(\ref{eq:baleqconc}) without bulk Soret effect and
scale the mode amplitudes according to
(\ref{eq:scalex})--(\ref{eq:scalev1v2}).
Then, the following model for the convection
in binary fluid mixtures is obtained:
\begin{mathletters}
\label{modell}
\begin{eqnarray}
\label{eq:modellx}
\tau {\bf\dot{{\bf X}}} & = & - \tilde{\sigma}\left[{\bf X}\ -\ {\bf Y}\ +\
      a\psi\left({\bf U_1}+\frac{1}{2}
      {\bf U_2}\right)\right]\\
\label{eq:modellz}
\tau \dot{Z} & = & -2\left( Z\ -\ {\bf X}\!\cdot\!{\bf Y}\right)\\
\label{eq:modelly}
\tau {\bf\dot{{\bf Y}}} & = & - {\bf Y}\ +\ {\bf X}\left(r-Z\right)\\
\label{eq:modellv1}
\tau \dot{V_1} & = & -\frac{6L}{5} V_1 - \frac{32L}{15} V_2 +
      {\bf X}\cdot\left({\bf U_1}+\,\frac{7}{3}{\bf U_2}\right)\\
\label{eq:modellv2}
\tau \dot{V_2} & = & -\frac{6L}{5} V_1 - \frac{24L}{5} V_2 +
      {\bf X}\cdot\left({\bf U_1}+\,4{\bf U_2}\right)\\
\label{eq:modellu1}
\tau {\bf\dot{{\bf U_1}}} & = & - r{\bf X} - \frac{L}{2}{\bf U_1} -
      \frac{5}{2}a\left( V_1-\frac{4}{9} V_2\right){\bf X}\\
\label{eq:modellu2}
\tau {\bf\dot{{\bf U_2}}} & = & - r{\bf X} - \frac{5L}{2}{\bf U_2} -
      \frac{10}{3}a\left( V_1-\frac{1}{6} V_2\right){\bf X}\ \ .
\end{eqnarray}
\end{mathletters}
It is an extension of the standard Lorenz model \cite{L63}.
The latter is contained in Eqs.~(\ref{eq:modellx})--(\ref{eq:modelly}) in
a form that is slightly modified due to a different
scaling and to the realistic no slip boundary conditions in our
approximation.

This model can be looked upon as a minimal one for convection in
binary liquid mixtures because it contains on the one hand the minimal
description of convection in a pure fluid (the Lorenz model) and on the
other hand a minimal extension for binary fluids. This extension is minimal
since the simple extension \cite{C86a,AL87,LLMN88,L90} of the Lorenz model
with only one linear mode ${\bf U}$ and one nonlinear mode $V$
leads to TW solutions with linear relations
between all pairs of the three quantities $v^2$
(the square of the phase velocity), $w^2$ (convective intensity),
and the Rayleigh number $r$. In Ref.~\cite{H93} it has been shown that such
pairwise linear relations result from {\em any\/} truncation of the
concentration field that is limited
to only one linear and one nonlinear mode. Clearly,
these pairwise linear relations between $v^2$, $w^2$, and $r$ are
incompatible with the topology of a backwards Hopf bifurcation followed
by a saddle node bifurcation into a branch of stable strongly nonlinear TWs.
Thus, our incorporation of a second linear mode ${\bf U_2}$ and its nonlinear
partner mode $V_2$ can be seen as a first non--trivial step in
a (systematic) extension that goes beyond earlier models
\cite{C86a,AL87,LLMN88,L90}.

Up to now, no few--mode model has described the bifurcation
topology of TWs adequately: The problem was not so much the backwards Hopf
bifurcation but rather the transition to strongly nonlinear convection, the
saddle node, and finally the merging of the TW solution branch
with the upper SOC solution branch. This failure of the earlier
approximations is due to an insufficient representation of the concentration
field: It has not been truncated directly but rather the combination
$\zeta = C - \psi T$ with the temperature field has been introduced in order
to fulfill the impermeable boundary condition {\em exactly\/}.
However, when using the combined field $\zeta$,
high mode representations in both, $\zeta$
as well as in $T$ are required as
explained in \cite{HL97a}. By enforcing the impermeability of
the plates only {\em in the lateral average\/} we
avoid these difficulties in our truncation (\ref{eq:Cansatz}).

As an aside we mention that within another minimal approach Knobloch and
Moore \cite{KM90} have deduced a model for free slip permeable boundary
conditions. They aimed at a correct, analytical
representation of the primary bifurcation and the involved modes
which is possible for idealized boundary conditions. However, their
model does not show TWs comparable with those seen in
experiments and related simulations.

\section{Results}
\label{sec:IV}

Here, we elucidate the SOC, TW, and SW solutions of our model.

\subsection{Stationary convection}
\label{sec:IVA}

\subsubsection{Bifurcation properties of SOC states}

In the case of SOC all time derivatives in (\ref{modell})
vanish so that 
${\bf X}\ \parallel\ {\bf Y}\ \parallel\ 
 {\bf U_1}\ \parallel\ {\bf U_2}$
holds in the complex plane for the amplitudes of the laterally varying
modes. Without loss of generality we may therefore
choose all modes to be real. After elimination of the temperature and
concentration modes the bifurcation diagram $r_{\rm SOC}({\bf X}^2)$ can be
calculated as
\begin{mathletters}
\label{eq:rSOCxqalles}
\begin{eqnarray}
\label{eq:rSOCxq}
r_{\rm SOC}({\bf X}^2) & \!\!\!\! = & \!\!\!\!
 \frac{\displaystyle 1+{\bf X}^2}
       {\displaystyle 1\!\!+\!\!\frac{11}{5}a\frac{\psi}{L}\,
  \frac{\textstyle \left(1+{\bf X}^2\right)\left[1+\frac{25}{99}a
  \left(\frac{{\bf X}}{L}\right)^2\right]}
   {\textstyle 1\!\!+\!\!\frac{185}{36}a\left(\frac{{\bf X}}{L}\right)^2
     \!\!+\!\!\frac{625}{648}a^2\left(\frac{{\bf X}}{L}\right)^4}}\\
\label{eq:rSOCxqapprox}
 & \!\!\!\! \stackrel{{\bf X}\gg L}{\longrightarrow} &
   \frac{1+{\bf X}^2}{1+\frac{72}{125}\frac{L\,\psi}{{\bf X}^2}
   (1+{\bf X}^2)}\ \ \ .
\end{eqnarray}
\end{mathletters}
The relation (\ref{eq:rSOCxqalles}) between $r$ and ${\bf X}^2$ can be
inverted, e.~g., graphically to obtain the standard bifurcation diagram of,
say, ${\bf X}^2$ vs.~$r$.
The stationary stability threshold 
$r_{\rm stat} = r_{\rm SOC}({\bf X}=0)$
of the quiescent heat conducting state follows to be
\begin{equation}
r_{\rm stat}  = \frac{1}{1+\frac{11}{5}a\frac{\psi}{L}}
  \simeq \frac{1}{1+1.527\frac{\psi}{L}}\ \ .
\label{eq:rstat}
\end{equation}
It agrees quite well with the result
$r_{\rm stat}\simeq \left(1+1.538\frac{\psi}{L}\right)^{-1}$
of Galerkin approximations \cite{L90,HL95} that fulfill the concentration
boundary condition exactly. 

As a first SOC property we can determine the type of the stationary
bifurcation out of the quiescent heat conducting state. For Soret couplings
smaller than
\begin{equation}
\psi_{\rm SOC}^t\ =\
 -\frac{4}{43\,a^2}
  \frac{L^3}
       {1+\frac{557}{774\,a}L+\frac{68}{129\,a}L^2+\frac{32}{215\,a^2}L^3}
\end{equation}
a subcritical bifurcation is observed:
$\partial r/\partial{\bf X}^2<0$. The scaling of $\psi_{\rm SOC}^t$ with
$L^3$ agrees with earlier free slip \cite{LL87} and no slip
predictions \cite{SZ93}. For subcritical bifurcations the saddle node is
found at
\begin{mathletters}
\begin{eqnarray}
r_{\rm SOC}^s
 & = & 1 + \frac{12}{5}\sqrt{\frac{2}{5}}\sqrt{-L\psi} + O(L\psi)\\
 & \simeq & 1 + 1.518 \sqrt{-L\psi}
\end{eqnarray}
\end{mathletters}
being in good agreement with the numerically determined result
$r_{\rm SOC}^s \simeq 1 + 1.636 \sqrt{-L\psi}$ \cite[Eq.~(4.1)]{HL97a}.

Eq.~(\ref{eq:rSOCxqapprox}) shows that for convective amplitudes ${\bf X}\gg L$
the Rayleigh number corresponding to a certain amplitude square ${\bf X}^2$
deviates from that of the pure fluid $r_{\psi=0}=1+{\bf X}^2$ only by terms
$\propto L\psi$.
This means that for convective amplitudes ${\bf X}^2\gg L|\psi|$ the
bifurcation diagrams of a mixture are the same as for a pure fluid. This
equality reflects the fact that strong convective mixing
in conjunction with diffusion equilibrates the concentration in the whole
fluid with the exception of narrow boundary layers
so that it does not influence the bifurcation
behavior any more: The stronger the mixing the smaller
the deviations from the pure fluid case.

\subsubsection{Fields}

In order to reconstruct the fields and with them the order parameters
such as the Nusselt number $N$ and the concentration variance $M$ 
(\ref{eq:Mdef}) we need the mode amplitudes
\begin{mathletters}
\begin{eqnarray}
{\bf Y} & = & \frac{r\,{\bf X}}{1+{\bf X}^2}\\
Z & = & \frac{r\,{\bf X}^2}{1+{\bf X}^2}\\
V_1 & = & -2\,r\,F_{\rm SOC}\,\left(\frac{{\bf X}}{L}\right)^2\,
  \left[1+\frac{25}{216}a\left(\frac{{\bf X}}{L}\right)^2\right]\\
V_2 & = & -\frac{1}{4}\,r\,F_{\rm SOC}\,\left(\frac{{\bf X}}{L}\right)^2\,
  \left[1-\frac{25}{18}a\left(\frac{{\bf X}}{L}\right)^2\right]\\
{\bf U_1} & = & -2\,r\,F_{\rm SOC}\,\left(\frac{{\bf X}}{L}\right)\,
  \left[1+\frac{5}{12}a\left(\frac{{\bf X}}{L}\right)^2\right]\\
{\bf U_2} & = & -\frac{2}{5}\,r\,F_{\rm SOC}\,\left(\frac{{\bf X}}{L}\right)\,
  \left[1-\frac{25}{18}a\left(\frac{{\bf X}}{L}\right)^2\right]
\end{eqnarray}
in the SOC fixed points. Here, we have introduced the quantity
\begin{equation}
F_{\rm SOC}\ =\ \left[1+\frac{185}{36}a\left(\frac{{\bf X}}{L}\right)^2
     +\frac{625}{648}a^2\left(\frac{{\bf X}}{L}\right)^4\right]^{-1}
\end{equation}
\end{mathletters}
for notational convenience. The square of $M$ is given by
\begin{eqnarray}
\label{eq:MSOC}
M^2_{\rm SOC}\ = 
 & \bigg[ 1 
 &   + 3.905\left(\frac{{\bf X}}{L}\right)^2
     + 2.224\left(\frac{{\bf X}}{L}\right)^4 \nonumber\\
 & & + 0.3040\left(\frac{{\bf X}}{L}\right)^6
     + 0.002072\left(\frac{{\bf X}}{L}\right)^8
 \bigg]\\
 \times & 
 \bigg[1 & + 3.566\left(\frac{{\bf X}}{L}\right)^2
           + 0.4645\left(\frac{{\bf X}}{L}\right)^4\bigg]^{-2}
 \nonumber
\end{eqnarray}
and the Nusselt number by
\begin{equation}
N_{\rm SOC}\ =\ 1 + \frac{25}{18}\,\frac{{\bf X}^2}{1+{\bf X}^2}\ \ .
\end{equation}

\subsubsection{Comparison with numerical results}

\begin{figure}[t]
\centerline{\psfig{figure=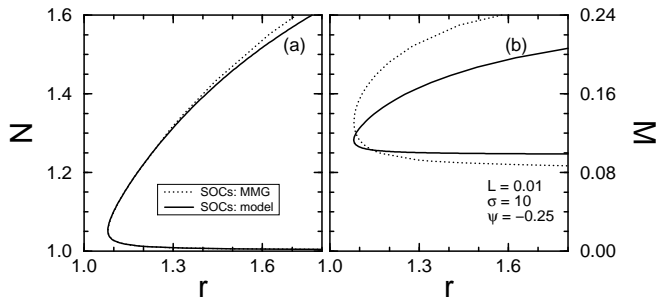,width=\linewidth,angle=270}\vspace*{3mm}}
\caption{SOC bifurcation diagrams of Nusselt number $N$ (a) and
         concentration variance $M$ (b) vs.~reduced Rayleigh number $r$.
         Exact (model) results are shown by dotted (solid) lines.}
\label{fig:BifVglSOC}
\end{figure}
Using these formulae we can compare the results of the model with "exact"
ones obtained by a MMG calculation \cite{HL97a}. This is
done in Fig.~\ref{fig:BifVglSOC} with the bifurcation diagrams of $N$ and $M$. 
Since the Nusselt number [Fig.~\ref{fig:BifVglSOC}(a)]
is determined by the well described temperature
field the Nusselt number of the model
deviates from the "exact" one maximally by $1$\%
(at $r=1.6$). The variance of the concentration field
$M$ in Fig.~\ref{fig:BifVglSOC}(b) shows that also the concentration field
is approximated reasonably well: the strong mixing
in stable SOCs with large velocity amplitudes
(upper branch of $N$ and lower branch of $M$) leading to nearly
equilibrated concentration distribution and nearly vanishing $M$
is reproduced with a relative error of about $15$\%. Since with
the special mode selection in the concentration field the model was
constructed to describe strongly nonlinear convection rather than weakly
nonlinear states it is not
surprising that $M$ in the unstable SOCs is reproduced only with an
accuracy of about $20$\% at $r=1.6$.

\begin{figure}[t]
\centerline{\psfig{figure=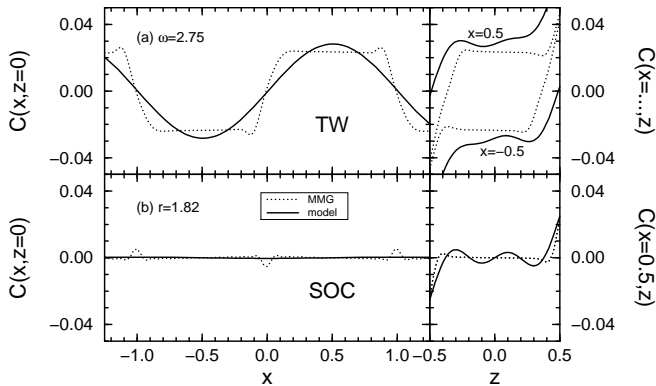,width=\linewidth,angle=270}\vspace*{3mm}}
\caption{Lateral (left column) and vertical (right column) concentration
         profiles of a TW with frequency $\omega=2.75$ (a) and of a SOC (b)
         at a Rayleigh number of $r=1.82$. The two line types compare model
         (solid lines) and "exact" results (dotted lines). Parameters are
         $L=0.01$, $\sigma=10$, and $\psi=-0.25$.}
\label{fig:Konzschnitt}
\end{figure}
Beyond these global order parameters we may also discuss the spatial
variations of the concentration field. In Fig.~\ref{fig:Konzschnitt}
we compare the SOC concentration field structure obatined from our model
with the "exact" one from a MMG scheme. To that end we show vertical
(right column) and horizontal (left column) profiles. Of course, the
model cannot describe the narrow concentration peaks in the lateral
direction which are due to the strong boundary layer phenomena caused by
the smallness of the ratio $L/w_{\rm max} = O(0.001)$. Nevertheless, the
model predicts that the concentration vanishes nearly all over the
convection cell. Also in the vertical profile we see a good agreement when
keeping in mind that the combination of only two modes, namely $c_{02}$
and $c_{04}$, can provide only a very rough approximation to a boundary
layer.

The quality of the approximation of the concentration can also be discussed
by its zeroth lateral Fourier mode. The model predicts
\begin{equation}
\label{eq:C0modellres}
C_0(z) = -\frac{\psi}{5}\left[z-\frac{1}{\pi}\sum_{n=1}^2
  (-1)^{n+1}\frac{\sin n(2\pi z)}{n}\right]\!+\!
  O\left(\frac{L}{{\bf X}}\right)^{\!\!2}\, .
\end{equation}
In the "exact" results \cite{BLKS95I,HL97a}, $C_0(z)$ is nearly zero in the
bulk of the fluid outside the boundary layers near the plates. It is
interesting to note that an extension of the series in the above expression
to $n=\infty$ would yield
\begin{equation}
\label{eq:reihe}
\frac{1}{\pi}\,\sum_{n=1}^\infty (-1)^{n+1}\,\frac{\sin n(2\pi z)}{n}\
=\ \frac{1}{\pi}\frac{2\pi z}{2}\ =\ z
\end{equation}
for values of $z \in ]-0.5,0.5[$ so that the linear term $z$ in the square
bracket of (\ref{eq:C0modellres}) is completely cancelled. Thus, 
the exact result --- $ C_0(z) \equiv 0 $ in the limit $L\rightarrow0$ ---
is reproduced in an optimal way, namely by
giving the exact results for those modes that the model contains. A
significantly improved description is possible only by using much more modes
because the contribution from higher modes in (\ref{eq:reihe})
decrease only $\propto 1/n$.

To summarize: stationary convection in binary liquid
mixtures is described for negative as well as for positive $\psi$
in a semi--quantitative way by our model.
Beyond topological details of the bifurcation diagrams even the
peculiar spatial structures of the concentration field can be explained.

\subsection{Traveling wave convection}
\label{sec:IVB}

\subsubsection{Bifurcation and scaling properties}

For the TW fixed points of the model (\ref{modell}) with a frequency $\omega$
we have to assume time dependences $ \propto e^{i\omega t} $
of the complex modes  ${\bf X}$, ${\bf Y}$,
${\bf U_1}$, and ${\bf U_2}$ because they are
the amplitudes of a lateral variation $ \propto e^{-ikx} $. So, positive
frequencies correspond for positive wave numbers $k$ to TWs traveling to
the right. The zeroth lateral modes $Z$, $V_1$, and $V_2$
are time independent in TWs. We separate the time dependence,
$e^{i\omega t}$, of the complex amplitude vectors by
\begin{equation}
\Big[{\bf X}(t) , {\bf Y}(t) , {\bf U_1}(t) , {\bf U_2}(t)\Big]
 = \Big[{\bf X} , {\bf Y} , {\bf U_1} , {\bf U_2}\Big]
   e^{i\omega t}
\end{equation}
and use henceforth the same symbols for the time independent prefactors.
Then, by choosing the temporal phase, ${\bf X}$ can be taken as real while 
${\bf Y},{\bf U_1},{\bf U_2} \in \mathbb{C}$. In addition,
$Z,V_1,V_2 \in \mathbb{R}$. Inserting these solution ansatzes into
(\ref{modell}) yields in order $O(L^0)$ the relations
\begin{mathletters}
\label{eq:TWerg}
\begin{eqnarray}
\label{eq:TWlsg1}
-\frac{1}{\psi} & = &
\frac{3}{2}a\frac{\tilde{\sigma}}{\tilde{\sigma}+1}
\frac{1\!+\!\Omega^2\!+\!{\bf X}^2}{\Omega^2}
\frac{1+\frac{175}{108}a\left(\frac{{\bf X}}{\Omega}\right)^2}
     {1\!+\!\frac{1025}{144}a\!\left(\frac{{\bf X}}{\Omega}\right)^2
       \!+\!\frac{15625}{10368}a^2\!\left(\frac{{\bf X}}{\Omega}\right)^4}
     \nonumber\\[-1mm]
 & & \\
\label{eq:TWlsg2}
r_{\rm TW} & = & 1 + \Omega^2 + {\bf X}^2
\end{eqnarray}
between frequency $\omega$, amplitude ${\bf X}$, and control parameter
$r_{\rm TW}$ of the TW solution. Here,
\begin{equation}
\Omega = \omega \tau\
\end{equation}
\end{mathletters}
has been introduced with $\tau=\frac{1}{2\pi^2}$ (\ref{eq:taudef}) being
the intrinsic time scale of the model (\ref{modell}).

Neglecting terms $O(L)$ in (\ref{eq:TWerg}) causes the TW to merge
at $\omega=0$ with the SOC solution branch of the pure fluid 
instead of with the SOC solution (\ref{eq:rSOCxqalles}) of the mixture.
Cancelling the terms of $O(L^2)$ is allowed for all states with
$L\ll\Omega$ or $L\ll{\bf X}$. This condition is fulfilled for all separation
ratios away from the co--dimension two (CT) point where Hopf
bifurcation {\em and\/} SOC--TW merging fall together and none of the
relations $L\ll\Omega$ or $L\ll{\bf X}$ can be fulfilled.
The TW fixed points of our model can be
calculated analytically without limiting to the order $O(L^0)$. Since the
formulae are lengthy they are not presented here. But they have
been used for the calculation of the phase diagram in
Fig.~\ref{fig:Phase} including the CT point. 

From (\ref{eq:TWlsg1}) one observes first of all
that TWs exist only for negative
Soret couplings $\psi<0$. This is in
line with the absence of TWs for $\psi>0$
in numerical simulations --- TWs for $\psi>0$ seen in the model
of Ref.~\cite{LLMN88} result from low--order truncation.
Relation (\ref{eq:TWlsg2}) allows to
determine the frequency $\Omega$ of a TW with a given velocity amplitude
${\bf X}$:
\begin{mathletters}
\begin{eqnarray}
r_{\rm TW}({\bf X}^2) & = & 1 + {\bf X}^2 + \Omega^2({\bf X}^2)\nonumber\\
 & = & r_{\psi=0}({\bf X}^2) + \Omega^2({\bf X}^2)\label{eq:freqskal}\\
\makebox[25mm]{or\hfill
$\frac{\omega^2({\bf X}^2)}{\omega_{\rm H}^2}$} & = &
 \frac{r_{\rm TW}({\bf X}^2) - r_{\psi=0}({\bf X}^2)}
      {r_{\rm osc}-1}\label{eq:univfreqskal}\ \ .
\end{eqnarray}
\end{mathletters}
Thus, the model predicts that the square of the frequency $\Omega$ of a TW
state with a velocity amplitude ${\bf X}$ is the distance in the
control parameter $r$ between the TW under consideration and the state of
the pure fluid with the same velocity amplitude. Hence, the TW frequency
is a direct measure of the distance of the system from the pure fluid,
i.~e., the influence of the concentration. 

Another equivalent interpretation of (\ref{eq:TWlsg2}) is that for a given
fixed $r$ the squared frequency of a TW with velocity amplitude
${\bf X}_{\rm TW}$,
\begin{equation}
\Omega^2 = {\bf X}_{\psi=0}^2 - {\bf X}_{\rm TW}^2\ \ ,
\end{equation}
is given by the difference between the pure fluid flow intensity,
${\bf X}_{\psi=0}^2 = r-1$, and the flow intensity ${\bf X}_{\rm TW}^2$ of
the TW in question. Thus, $\Omega^2$ measures also the "vertical" distance 
in the bifurcation diagrams of ${\bf X}^2$ vs.~$r$ between the $\psi=0$
pure fluid SOC solution and the TW solution in the mixture.

\begin{figure}[t]
\centerline{\psfig{figure=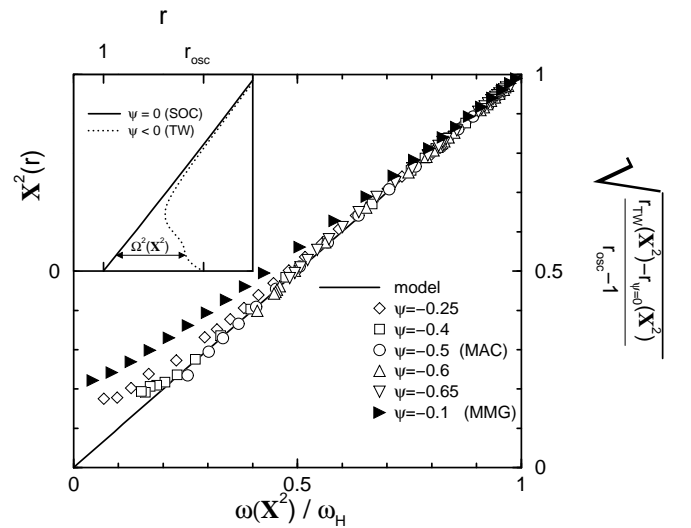,width=\linewidth,angle=270}\vspace*{3mm}}
\caption{Universal scaling relation connecting TW convective velocity
         amplitude ${\bf X} \propto w_{\rm max}$ with its frequency
         $\omega$. The model prediction (\ref{eq:univfreqskal}) is
         the identity (solid line). Numerical, finite differences (MAC)
         data are shown by open symbols for $\psi \in
         [-0.25,-0.65]$ and MMG data by filled symbols ($\psi=-0.1$). The
         symbols cover in each case the whole bifurcation branch, i~e.,
         stable as well as unstable TW states. The inset serves as a
         schematic explanation of the scaling relation. $L$ and $\sigma$
         were fixed to values of $0.01$ and $10$, respectively. But the
         scaling relation should not depend on them as long as
         $\sigma \gtrsim 1$ and $L \ll 1$.}
\label{fig:UnivFreqSkal}
\end{figure}
Eq.~(\ref{eq:freqskal}) has an
explicit dependence on the Soret coupling strength $\psi$ since
$\Omega$ varies between $0$ and the scaled Hopf frequency
$\Omega_{\rm H}$. This $\psi$--dependence is cancelled by scaling 
(\ref{eq:freqskal}) with the Hopf frequency so that the left hand side
of (\ref{eq:univfreqskal}) varies for all $\psi$ between $0$ and $1$. 
Thus, Eq.~(\ref{eq:univfreqskal}) is a universal scaling relation for TW
frequencies resulting from our model for small $L$. 
In Fig.~\ref{fig:UnivFreqSkal} this
prediction of the model is compared with 
numerical results for $\psi \in [-0.65,-0.25]$
obtained by a finite difference scheme and a MMG scheme \cite{HBL97}.
For all these Soret couplings the scaling relation is
confirmed by the numerical results. Only in the case of small frequencies
deviations are observed. They are due to the fact that TWs with
small frequencies do not approach the SOC states of the pure fluid
as implied by (\ref{eq:TWlsg2}) but rather the SOC states
of the mixture. These two stationary
states differ in the Rayleigh number $r$ by $O(L)$ according 
(\ref{eq:rSOCxqapprox}) so that deviations of the order
$O(\sqrt{L}) \simeq 0.1$ can be expected on the ordinate of
Fig.~\ref{fig:UnivFreqSkal}. This deviation from the scaling relation
becomes more obvious for weaker Soret couplings like, e.~g., $\psi=-0.1$ as
shown by the filled triangles. These states have been computed by a 
MMG scheme \cite{HL97a}. For small Soret couplings the merging point of the
SOC and TW branches is in a regime
of small velocity amplitudes, i.~e., in that part of the bifurcation
diagram where the differences between SOC states of pure and binary
mixed fluids become more and more evident.

It should be noticed that the scaling relation holds for stable as well as
for unstable TWs. Furthermore, it is independent of the
bifurcation topology, i.~e., it applies to
the form shown in the inset of Fig.~\ref{fig:UnivFreqSkal},
i.~e., for $\psi \gtrsim -0.4$ as well as to a topology with
a bistability of slow and fast TWs ($\psi\lesssim -0.4$) \cite{HBL97}.

Bifurcation properties are best discussed by introducing the ratio
\begin{equation}
\label{eq:chidef}
\chi\ = \frac{w_{max}}{v}\ =\
           \frac{5\,\pi}{8}\,\frac{|{\bf X}|}{\Omega}
\end{equation}          
of convective and phase velocity of the TW state. Then, the Rayleigh number
$r_{\rm TW}(\chi^2)$ of a TW with a velocity ratio $\chi$ can be
written in order $O(L^0)$ as
\begin{equation}
\label{eq:rTWvonchi}
r_{\rm TW}(\chi^2) =
 \frac{\displaystyle 1}{\displaystyle 1 + 
  \frac{\displaystyle 27\,\pi^2}{\displaystyle 256}\,\psi\,
  \frac{\displaystyle \tilde{\sigma}}{\displaystyle 1+\tilde{\sigma}}\,
  \frac{\rule[-2mm]{0mm}{3mm}\textstyle \left(1\!\!+\!\!\frac{64}{25\,\pi^2}\chi^2\right)
                                        \left(1\!\!+\!\!\frac{7}{24}\chi^2\right)}
       {\rule{0mm}{4mm}\textstyle 1+\frac{41}{32}\chi^2
                                   +\frac{25}{512}\chi^4}}\ \ .
\end{equation}
The oscillatory stability threshold of the basic state follows for $\chi=0$
as 
\begin{mathletters}
\begin{equation}
r_{\rm osc}\ =\ r_{\rm TW}(0)\ =\
   \frac{1}{1+\frac{27\pi^2}{256}
            \frac{\tilde{\sigma}}{1+\tilde{\sigma}}\psi}
\end{equation}
and (\ref{eq:TWlsg2}) yields the Hopf frequency
\begin{equation}
\omega_{\rm H}^2\ =\
 \frac{-\frac{27\,\pi^6}{64}\,\psi}
      {\frac{1+\tilde{\sigma}}{\tilde{\sigma}}\,+\,
       \frac{27\,\pi^2}{256}\psi}\ \simeq\
 \frac{- 405.6\,\psi}{\frac{1+\tilde{\sigma}}{\tilde{\sigma}}\,+\,
       1.041\,\psi}\ \ . 
\end{equation}
\end{mathletters}
The calculation of the CT point requires the consideration
of the Lewis number dependence. If this is done one obtains
\begin{equation}
\psi^{\rm CT}\ =\ -\frac{1600}{459\,\pi^2}\,
                   \frac{\tilde{\sigma}+1}
                        {\tilde{\sigma}+\frac{55}{102}L}\,\,L^2
\end{equation}
as the separation ratio at the CT point for given Lewis--
and Prandtl number and fixed wave number $k=\pi$.
In ethanol--water mixtures with
$L = 0.01$ and $\sigma = 10$, this yields
$\psi^{\rm CT}=-3.714\cdot 10^{-5}$ being in very good agreement with the
numerically \cite{SZ93} determined value of $-3.526\cdot 10^{-5}$. 

The other limit in $\chi$, namely $\chi\rightarrow\infty$ or
$\omega \rightarrow 0$, gives the Rayleigh number of the SOC--TW transition:
\begin{equation}
\label{eq:rstern}
r^* = \lim_{\chi\rightarrow\infty}r_{TW}(\chi^2)\ =\
         \frac{1}{1+\frac{1008}{625}
                    \frac{\tilde{\sigma}}{1+\tilde{\sigma}}\psi}\ \ .
\end{equation}
However, one should keep in mind that the SOC--TW transition at $r^*$
with the transfer of stability from an SOC to a TW when reducing $r$ is
related to an instability of the SOC concentration boundary layer
\cite{BPS89}. These boundary
layers are caused by the smallness of the Lewis number $L$, a limit which
is not systematically incorporated in the model under consideration. The
model's main objective is a description of strongly nonlinear TW
convection which has for finite TW frequencies a definite limit for small
$L$. Thus, one should not expect a correct reproduction of the Lewis number
dependence of the SOC--TW transition from a model with modes that do not
resolve the boundary layer structure in detail.
However, the dependence on the two other fluid parameters, namely
separation ratio $\psi$ and Prandtl number $\sigma$ is given in a
qualitatively correct way: strong increase of $r^*$ with stronger negative
Soret coupling and saturation with increasing $\sigma$ as can be
seen by comparing the formula (\ref{eq:rstern}) with the numerical results
in \cite[Fig.~9(b) and 15(b)]{BLKS95I}. The SOC--TW transition points which
have been plotted therein are also affected with a certain error bar since the
spatial resolution of the used numerical method was not fine enough to
capture the whole boundary layer phenomena. However, their qualitative fluid
parameter dependence has to be regarded as correct.

The analytical form (\ref{eq:rTWvonchi}) for the TW bifurcation diagram
allows a simple determination of the TW saddle node bifurcation, namely as
the minimum of $r_{\rm TW}(\chi^2)$. It is given by
\begin{equation}
 r_{TW}^s \simeq r_{TW}(2.8687) \simeq
   \frac{1}{1+0.6567\frac{\tilde{\sigma}}{1+\tilde{\sigma}}\psi}\ \ .
\end{equation}

\subsubsection{Fields}

The calculation of Nusselt number $N$,
concentration variance $M$, and concentration contrast between the two
plates requires the computation of the temperature and concentration field in
the TW fixed points. Their mode amplitudes in the
TW fixed points can be expressed by
\begin{mathletters}
\begin{eqnarray}
\label{eq:TWmodenlsg}
V_1 & = & -\frac{32}{5\,\pi^2}\,\chi^2\,r\,F_{\rm TW}\,
 \left(1+\frac{5}{192}\chi^2\right)\nonumber\\
V_2 & = & -\frac{4}{\pi^2}\,\chi^2\,r\,F_{\rm TW}\,
 \left(1-\frac{1}{16}\chi^2\right)\nonumber\\
{\bf U_1} & = & -\frac{\bf X}{\Omega}\left(\frac{L}{2\Omega}-i\right) \,r\,F_{\rm TW}\,
 \left(1+\frac{15}{32}\chi^2\right)\\
{\bf U_2} & = & -\frac{\bf X}{\Omega}\left(\frac{5L}{2\Omega}-i\right) \,r\,F_{\rm TW}\,
 \left(1-\frac{1}{16}\chi^2\right)\nonumber\\
{\bf Y}   & = & {\bf X}\,(1-i\Omega)\nonumber\\
Z & = & {\bf X}^2\nonumber
\end{eqnarray}
where we have introduced the quantity
\begin{equation}
\label{eq:ftwdef}
F_{\rm TW} = \left(1+\frac{41}{32}\chi^2
                    +\frac{25}{512}\chi^4\right)^{-1}\ \ .
\end{equation}
\end{mathletters}
In (\ref{eq:TWmodenlsg}) all quantities except the real parts of ${\bf U_1}$
and ${\bf U_2}$ are evaluated in order $L^0$ with ${\bf X}$ and $\Omega$
taken from (\ref{eq:TWerg}).

\subsubsection{Small amplitude expansions}

Before discussing the order parameters themselves we should like to show
that they {\em cannot\/} be expanded as power series in the distance
\begin{equation}
\label{eq:epsilon}
\epsilon = \frac{r_{\rm TW}(\chi^2)-r_{\rm osc}}{r_{\rm osc}}
\end{equation}
from the onset of convection
up to values where strongly nonlinear TW convection is
observed. To see this, let us rewrite
Eq.~(\ref{eq:epsilon}) by using (\ref{eq:rTWvonchi}) to display the
relation between $\epsilon$ and $\chi^2$ explicitly
\begin{equation}
\label{eq:epsilonvonchi}
\epsilon = \frac{a_1\chi^2\,(1+a_2\chi^2)}{1+b_1\chi^2+b_2\chi^4}
         = \frac{a_1\chi^2\,(1+a_2\chi^2)}
                {\left(1-\frac{\chi^2}{b_1'}\right)
                 \left(1-\frac{\chi^2}{b_2'}\right)}\ \ .
\end{equation}
Here, $a_{1,2}$, $b_{1,2}$, and $b_{1,2}'$ are amplitude independent real
numbers. Note in particular that the functional relation between control
parameter $\epsilon$ and reduced order parameter $\chi=w_{\rm max}/v$ is
given by a rational function. A similar relation has also been found from a
fit to MMG and finite difference numerical results \cite{HBL97}.
The radius of convergence of a small amplitude expansion of
(\ref{eq:epsilonvonchi}) in powers of $\chi^2$ is given by
$$ \chi_c^2\ =\ \min_{i=1,2}|b_i'|\ \ . $$
This quantity depends on $\psi$ and $\sigma$ so that for all negative values
of $\psi$ (TWs are observed only for $\psi<0$ according to (\ref{eq:TWlsg1}))
$$ \chi_c^2\ \leq \frac{328}{25}
 \left(1-\sqrt{\frac{1481}{1681}}\right)\ \simeq\ 0.8052 $$
which is the absolute value of that node of $F_{\rm TW}^{-1}(\chi^2)$
(\ref{eq:ftwdef})
with the smallest absolute value. Then, the radius of convergence is
calculated in the variable $w_{\rm max}/v$ as
\begin{equation}
\label{eq:konvrad}
\left(\frac{w_{max}}{v}\right)_c\ =\ \chi_c\ \simeq\ 0.8973\ \ ,
\end{equation}
i.~e., near that point in the bifurcation diagram which has been identified
in \cite{HBL97} as the transition between weakly and strongly nonlinear
convection. This point, namely $w_{\rm max} \simeq v$, where areas of
closed streamlines first occur, has also been identified as the radius of
convergence for a small amplitude power series expansion of different
order parameters (see Ref.~\cite{HBL97}).

Thus, our model supports the notion that experimentally observed TW convection
in binary liquid mixtures {\em cannot\/} be described
by weakly nonlinear models as, e.~g., complex Ginzburg--Landau amplitude
equations (GL) including various ad hoc quintic extensions that have been
proposed. Being used out of their validity range they cannot be trusted to
reproduce, e.~g., the relations between frequency $\omega$, mixing $M$,
flow intensity $w_{\rm max}^2$ or Nusselt number $N$, and the thermal
driving $r$. Typically already the simpler relation between $w_{\rm max}^2$
or $N$ and $r$ is wrong on the upper TW branch --- not to mention the more
sensitive relations between $\omega$, $M$, and $r$.
Also results for LTWs based on this approach \cite{CGLEs} have to be
questioned: The spatiotemporal field properties under the envelope being
closely related to those in extended TWs are not captured properly. The
main drawback of these GL approaches is the insufficient representation of
the role of the concentration field in these strongly nonlinear states.

A first step towards a better incorporation of the concentration field into
the GL framework was the introduction of a long wavelength concentration 
mode with characteristic time scale $\propto L$ by Riecke \cite{Riecke}.
The approximation is that $L$ is of the same order as the distance from
onset which, however, does not apply to all experimental LTWs. Additionally,
those parameters ($h_1$ and $h_3$ in Ref.~\cite{Riecke}) that could 
immediately lead in extended TW
states without large scale lateral variation to a finite mean concentration
mode (possibly at the expense of stabilizing terms of unphysical fifth
order) have been dropped in the LTW calculations \cite{Riecke}. While this
approach in its present form does not seem to generate the spatiotemporal
field structure of TWs under the LTW envelope it is a promising step
forwards. Incorporation of impermeable boundary conditions and separation
of diffusive ($L$) and critical ($\epsilon$) time scales and thus the
incorporation of an additional concentration mode seems necessary to
guarantee the aforementioned relations between $\omega$, $M$, $N$, and $r$.

\subsubsection{Comparison with numerical results}

\begin{figure}[t]
\centerline{\psfig{figure=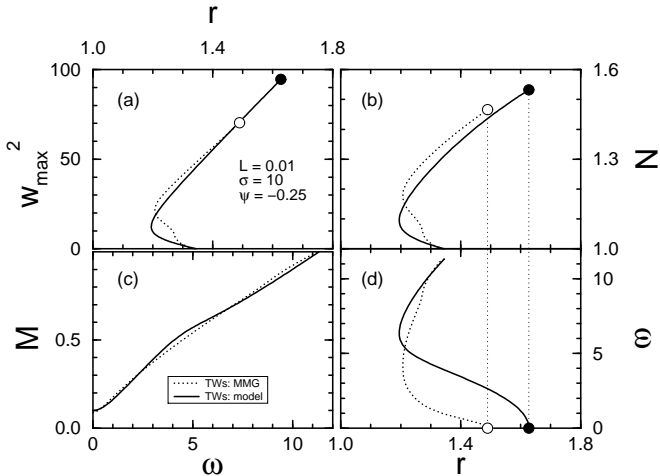,width=\linewidth,angle=270}\vspace*{3mm}}
\caption{TW bifurcation diagrams of the square of the convective velocity
         amplitude $w_{\rm max}^2$ (a), the Nusselt number $N$ (b), and
         the frequency $\omega$ (d), each vs~$r$. In (c) the reduced
         concentration variance $M$ is plotted vs.~frequency.}
\label{fig:BifVglTW}
\end{figure}

For a quantitative comparison with numerical MMG results \cite{HL97a}
we present in Fig.~\ref{fig:BifVglTW}
bifurcation diagrams of the square of convective amplitude (a), the Nusselt
number (b), the TW frequency (d), and the concentration variance as a
function of the frequency (c). They show
that all characteristic features of the TW
bifurcation scenario are captured by the model: subcritical Hopf
bifurcation, saddle node bifurcation, stable upper branch of strongly
nonlinear TWs, merging of TWs and SOCs at $\omega=0$ in the
strongly nonlinear regime, and drastic reduction
of the concentration contrast with decreasing frequency. 
As an aside we mention that a model using the same number of modes,
but numerically determined ones is similarly successful \cite{HL97b}.
This has to be contrasted with {\em earlier analytical\/} few mode
approximations for TWs in binary mixtures which reproduced only the
backwards Hopf bifurcation.
 
The linear and weakly nonlinear bifurcation
properties, i.~e., the onset of convection and the initial slope, are
modelled by our approximation with high accuracy. The
characteristic bump in the "exact" bifurcation diagrams of
Fig.~\ref{fig:BifVglTW}(a), (b), and (d) that occurs at
the transition $\chi \simeq 1$ from
weakly to strongly nonlinear convection is not reproduced by our model.
The absence of this fine structure in the bifurcation diagram of our model
is due to neglecting Fourier modes higher than the first
lateral one in the concentration. We have explicitly checked this by
determining the bifurcation diagrams for fields restricted to their zeroth
and first lateral Fourier modes but with full vertical resolution. This is
also clear from studying the changes in the concentration at that point:
Starting at the onset, only harmonic lateral variation is observed up to a
velocity ratio $\chi = w_{\rm max}/v\ \simeq\ 1$.
For larger $\chi$, the harmonic
profile gets more and more deformed by the occurrence of
plateaus \cite[Fig.~2]{HBL97}. They reflect homogenized concentration
distributions in the regions of closed streamlines. The description of this
equilibration requires higher lateral Fourier modes.

The next characteristic point is the position of the TW saddle node
bifurcation: As a consequence of the dropping of higher lateral Fourier modes
and the above discussed implications, the saddle lies at too high frequencies
or too low amplitudes, i.~e., in a too weakly nonlinear part of the bifurcation
diagram. On the upper (lower) branch of the
$N$ vs.~$r$ ($\omega$ vs.~$r$) curve the
TW frequeny of our model has too high values [Fig.~\ref{fig:BifVglTW}(d)]
and the zero frequency TW end point at $r^*$ lies above the "exact" one.
This is caused by the fact that slow TWs are boundary layer dominated.
This feature is not fully reflected in our model.

Since the concentration changes significantly with frequency it is
appropriate to discuss the relation between concentration variance $M$ and
TW frequency $\omega$ [Fig.~\ref{fig:BifVglTW}(c)] 
rather than the relation between $M$ and $r$ thereby eliminating partly the
errors in our $\omega$ vs.~$r$ curve of Fig.~\ref{fig:BifVglTW}(d). The
prediction of the model for $M(\omega)$ agrees very well with the "exact"
curve in Fig.~\ref{fig:BifVglTW}(c).
This is once more a hint that the concentration field is globally
treated in an adequate manner. Additionally, the relation between $M$ and
$\omega$ is a second universal,
$\psi$--independent, scaling relation when scaling the
frequency with its value at the Hopf bifurcation. This is done in
\cite[Fig.~5b]{HBL97} with numerical data.

The spatial variation of the concentration field in a TW is shown in
Fig.~\ref{fig:Konzschnitt}(a). To measure the quality of the model
we compare its results with numerically obtained fields. As described
above, it is convenient to select for this procedure two TWs with the same
frequency but different Rayleigh numbers. The "effective" value of the
harmonic lateral profile [left part of Fig.~\ref{fig:Konzschnitt}(a)]
in the model corresponds well with the "exact" plateau--like
concentration distribution. In the vertical profile [right part of
Fig.~\ref{fig:Konzschnitt}(a)], even a slight building up of a plateau
can be observed. Its mean height is approximated by the height of
the lateral profile explaining the differences in the heights of the
vertical plateaus. The strong variation of the model concentration along
the plates is an artefact of only approximately fulfilling the
impermeability of the plates.

\begin{figure}[t]
\centerline{\psfig{figure=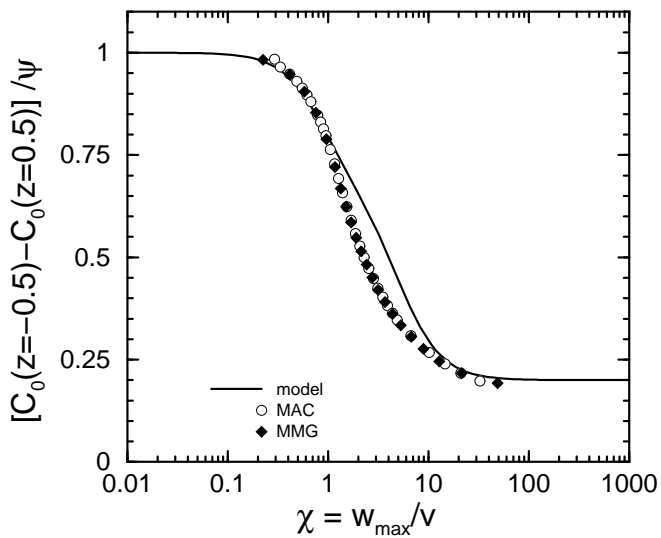,width=\linewidth}\vspace*{3mm}}
\caption{Concentration contrast in a TW state between the plates as a function
         of the velocity ratio $\chi=w_{\rm max}/v$ from the
         model (solid line) and from numerical computations
         (finite differences: open circles, many mode Galerkin: filled
         lozenges). Parameters are $L=0.01$, $\sigma=10$, and $\psi=-0.25$.}
\label{fig:Cplatte}
\end{figure}

In the actual TW states, the concentration
at the plates is nearly constant so that also the contrast
between them is nearly
constant. Thus, an appropriate quantity to compare is the laterally averaged
concentration contrast at the two plates. Our model predicts that
\begin{eqnarray}
\label{eq:cplattetw}
C_0(z=\frac{1}{2})-C_0(z=-\frac{1}{2})
 = 2 C_0(z=\frac{1}{2}) \text{\hspace{-50mm}} 
 \nonumber\\
 & = & -\psi
 \left(1\!\! -\!\! \frac{7}{16}\chi^2
 \frac{\rule[-2mm]{0mm}{3mm} 1+\frac{5}{56}\chi^2}
      {\rule{0mm}{4mm} 1\!\!+\!\!\frac{41}{32}\chi^2
                        \!\!+\!\!\frac{25}{512}\chi^4}
 \right)\\
 & \stackrel{\omega\rightarrow 0}{\longrightarrow} & -\frac{\psi}{5}\nonumber
\end{eqnarray}
depends only on the velocity ratio $w_{\rm max}/v$. This relation is
checked in Fig.~\ref{fig:Cplatte} by comparing the model prediction with
numerical finite differences and MMG results. The two
limits, namely the basic state with $w_{\rm max} \rightarrow 0$ and the
stationary state $v \rightarrow 0$ ($\chi\rightarrow\infty$)
are reproduced very well by our
approximation. The transition between them takes place for
$\chi \in [1,10]$.

\subsection{Standing wave convection}
\label{sec:IVC}

Besides SOC and TW convection there exists a third
type of convection that bifurcates out of the ground state: It
is standing wave convection occurring
at a Hopf bifurcation simultaneously
with TW convection. As long as linear states are considered
SWs are just a linear superposition of a
right and left traveling wave. The
selection of the nonlinear pattern leading to either a TW or SW
is governed by a general principle \cite{Kn86b}: 
stable solutions exist directly at a Hopf bifurcation only if
both SW {\em and\/} TW bifurcate supercritically. Then,
the pattern with the largest initial slope
of the amplitude is selected. Using
this principle and the numerically obtained initial slopes of \cite{SZ93}
it can be inferred that in all liquid mixtures TWs are stable when
bifurcating supercritically, i.~e., for $\psi$ larger than the tricritical
value $\psi_{\rm TW}^t \propto - L^2$.
Thus SWs cannot be observed
directly in experiments with liquid mixtures. Their investigation requires
to stabilize that unstable fixed point and 
possibly destabilizing the stationary
fixed point having the same symmetries. Nevertheless, SWs represent a
generic convection pattern in binary fluid mixtures. They occur in
particular as transients in the evolution of a strongly nonlinear TW out of
the unstable, supercritically heated ground state.

Up to now, only weakly nonlinear properties as the initial slope of SWs
have been discussed for binary fluid mixtures with
experimentally realistic, i.~e., impermeable boundary conditions. Since our
model (\ref{modell}) has described both SOCs and TWs in an adequate way we
think it is worthwhile to equally investigate
the bifurcation properties and time dependence of SWs in its framework.
The computation of the SW fixed points using the full field equations is a
problem which has not been solved so far. 

\subsubsection{Solution procedure}

In SWs the phases of the complex modes
${\bf X}$, ${\bf Y}$,
${\bf U_1}$, and ${\bf U_2}$
are time independent so that the time derivative is parallel to the mode
itself, e.~g., ${\bf X}\parallel\dot{{\bf X}}$.
Via the model (\ref{modell}) this leads to
$ {\bf X} \parallel {\bf Y} \parallel
 {\bf U_1} \parallel {\bf U_2}$,
i.~e., all modes may be chosen real without restriction of generality.

In the linear modes ${\bf X}$, ${\bf Y}$, ${\bf U_1}$, and ${\bf U_2}$
only odd multiples of the basic frequency $\omega$ of the SW occur whereas
in the nonlinear modes $Z$, $V_1$, and $V_2$ only
even multiples exist. This allows to expand the time dependence of the mode
amplitudes in the following way:
\begin{mathletters}
\label{eq:SWansatz}
\begin{eqnarray}
\label{eq:SWansatz1}
\makebox[50mm]{$
\left[{\bf X}(t),{\bf Y}(t),{\bf U_1}(t),
      {\bf U_2}(t)\right]$\hfill}\hspace{-46mm}
 & &\nonumber\\
 & = &
 \!\!\!\!\sum_{n=1,3,5,...}\!\!\!\!\!
 \left[{\bf X}^{(n)},\!{\bf Y}^{(n)},\!{\bf U_1}^{(n)},\!
       {\bf U_2}^{(n)}\right]
 e^{in\omega t} + \text{c.c.}\\
\label{eq:SWansatz2}
\makebox[50mm]{$
\left[Z(t),V_1(t),V_2(t)\right]$\hfill}\hspace{-46mm}
 & &\nonumber\\
 & = &
 \!\!\!\!\sum_{n=0,2,4,...}\!\!\!\!\!
 \left[Z^{(n)},V_1^{(n)},V_2^{(n)}\right] e^{in\omega t} + \text{c.c.}
\end{eqnarray}
\end{mathletters}
These series can be inserted in the model (\ref{modell}), the mode amplitude
${\bf X}^{(1)}$ can be chosen real by the arbitraryness of a common phase
in time, and the resulting nonlinear system of algebraic equations can
be solved in the unknown variables
$\{\omega,{\bf X}^{(1)}, \mbox{Re}\,{\bf X}^{(3)},
 \mbox{Im}\,{\bf X}^{(3)}, ...\}$.

\subsubsection{Bifurcation diagram and temporal behavior}

\begin{figure}[t]
\centerline{\psfig{figure=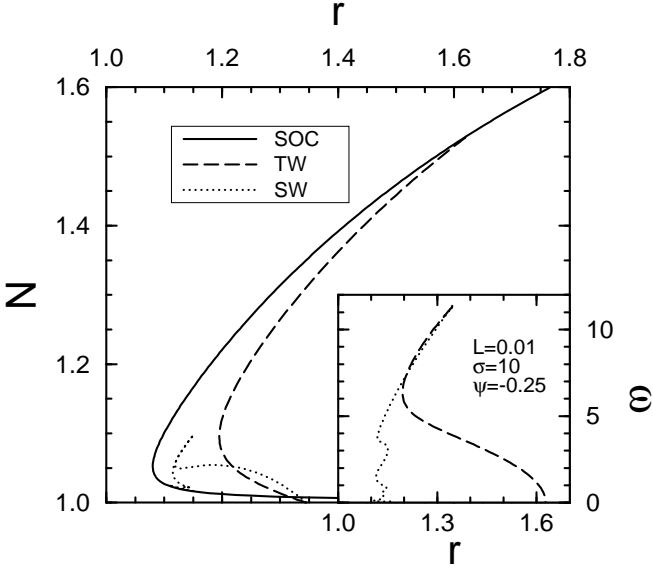,width=\linewidth}\vspace*{3mm}}
\caption{Complete bifurcation diagram of the Nusselt number (time averaged
         for SWs) of all types of convection connected
         with the ground by a primary bifurction: stationary convection
         (SOC), traveling (TW) and standing (SW) waves. The inset shows the
         frequency bifurcation diagrams of TWs and SWs.}
\label{fig:SWbif}
\end{figure}

\begin{figure}[t]
\centerline{\psfig{figure=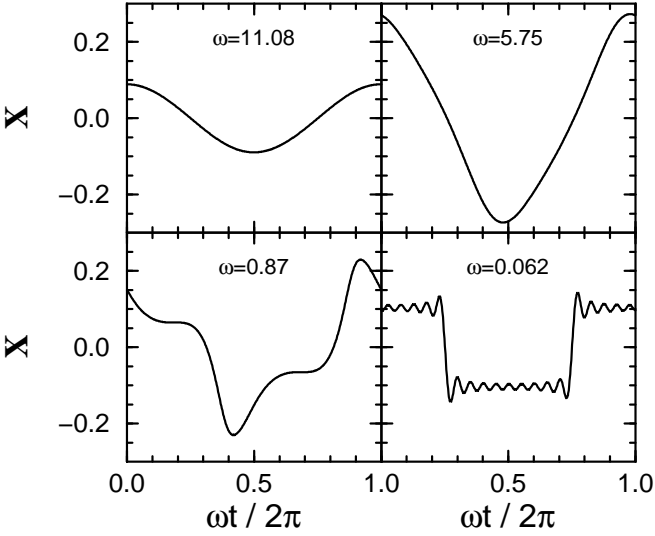,width=\linewidth}\vspace*{3mm}}
\caption{Time dependence of the velocity field amplitude {\bf X} over one
         oscillation period in SWs with four different
         frequencies. Parameters are $L=0.01$, $\sigma=10$, and $\psi=-0.25$.}
\label{fig:SWzeit}
\end{figure}

In Fig.~\ref{fig:SWbif} a bifurcation diagram
of the time averaged Nusselt number
in SWs (dotted line) is given for the standard ethanol--water fluid
parameters together with the already discussed diagrams for SOCs (solid
line) and TWs (dashed line). The inset shows the frequency of SW and TW.
At the oscillatory onset of convection ($r_{osc} \simeq 1.345$) both TWs
and SWs bifurcate subcritically as it has been predicted by \cite{SZ93}. With
decreasing frequency a weak maximum in the time averaged Nusselt number of
the SWs is observed ($r \simeq 1.19$) before a saddle node occurs at
$r \simeq 1.12$. Then, it increases with further decreasing frequency up to
a absolute maximum of about $1.1$ before approaching the SOC branch. The SW
branch has to be connected with the SOC branch since a SW with frequency $0$
is a SOC. The exact merging point of SWs and SOCs cannot be determined by
means of (\ref{eq:SWansatz}) since with decreasing frequency the
time dependence of the modes gets more and more anharmonic so that in 
the Fourier ansatz (\ref{eq:SWansatz}) more and more
modes have to be included. The evolution of this anharmonicity
with decreasing frequency is indicated in Fig.~\ref{fig:SWzeit} where the
time dependence of the velocity field mode ${\bf X}$ within one period of
oscillation is plotted.

For the computation of the SWs shown in Fig.~\ref{fig:SWbif} and
Fig.~\ref{fig:SWzeit} temporal Fourier modes up to $10\omega$ have been used
and compared with states calculated with modes up to
$20\omega$. Between $\omega_{\rm H}$ and $\omega = 0.4$ the results of 
both time resolutions agree in Nusselt and Rayleigh number better than
$1$\% for fixed frequency. However, states with $\omega < 0.4$ shown in 
Fig.~\ref{fig:SWbif} and Fig.~\ref{fig:SWzeit} may be erroneous.

\subsection{Phase diagram}
\begin{figure}[t]
\centerline{\psfig{figure=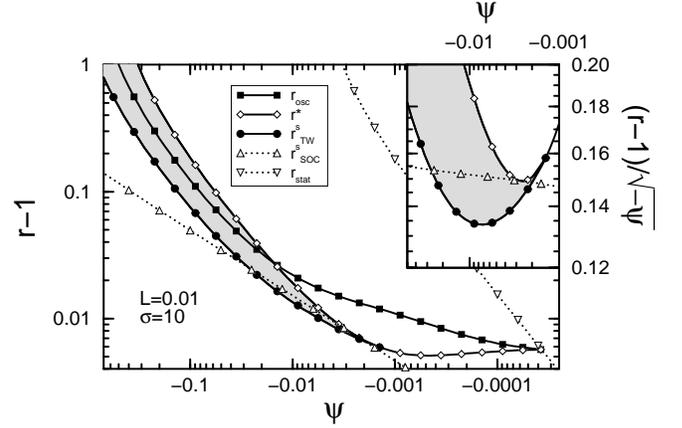,width=\linewidth}\vspace*{3mm}}
\caption{Phase diagram in mixtures with $L=0.01$
         and $\sigma=10$ in a double logarithmical plot $r-1$
         {\em vs.\/}~$\psi$.
         SOC properties are shown by dotted lines (saddle node $r_{\rm
         SOC}^s$: open triangles up, bifurcation $r_{\rm stat}$: open
         triangles down). Solid lines correspond to TWs
         (Hopf bifurcation $r_{\rm osc}$: filled
         squares, SOC--TW--transition $r^*$: open lozenges, saddle node
         $r_{\rm TW}^s$: filled circles). Stable TWs exist within the
         shaded area.}
\label{fig:Phase}
\end{figure}
The dependence of the bifurcation topology of SOCs and TWs
on the separation ratio is summarized
in the phase diagram of Fig.~\ref{fig:Phase} which can be compared directly
with the "exact" results described in \cite[Fig.~7]{HL97a}. The
characteristic points of the SOC bifurcation diagrams, namely the
stationary onset $r_{\rm stat}$ and the saddle
node $r_{\rm SOC}^{s}$ agree very well with the "exact" results. The same
holds without restriction for the oscillatory onset of convection
$r_{\rm osc}$ and the TW saddle node bifurcation $r_{\rm TW}^s$. The
$\psi$--dependence of the SOC--TW transition point $r^*$, i.~e., the
strong increase with increasing the negative coupling strength, is only
reproduced qualitatively: $r^*(\psi)$ runs too flat in the interval
$[-0.3,-0.01]$ so that the point where $r^* \equiv r_{\rm osc}$ is given as
$\psi \simeq -0.014$ instead of $-0.14$. This is due to the neglection
of the influence of boundary layers in the concentration field. The same is
true for the merging of $r^*$ with the TW saddle node, i.~e., that separation
ratio beyond which no stable upper TW branch can be observed: $\psi \simeq
-0.001$ instead of $-0.008$. Nevertheless, the model predicts
a $\psi$--interval where the TW saddle node can be seen below the SOC saddle
node, as it has been discussed in the framework of the "exact" results in
\cite[Fig.~5d]{HL97a}. Furthermore, the tricritical SOC--TW transition
\cite[Fig.~5e]{HL97a} at which 
the TW branch merges vertically with the SOC branch,
is predicted to take place on the unstable SOC branch, too.
For more negative $\psi$ a TW saddle appears. Then, 
stabilization of TWs at that saddle {\em and\/} subsequent
destabilization, probably towards modulated TWs,
occurs before the now unstable TWs end on the unstable lower SOC branch 
\cite[Fig.~6]{HL97a}.

\section{Conclusion}

We have investigated roll like 2D convection in binary liquid
mixtures with negative Soret coupling. Then, three types
of extended convective states occur which
are connected via a primary bifurcation to the quiescent heat conducting
state: stationary convection (SOC), traveling waves (TW), and standing
waves (SW). One objective of our paper was to derive a model
describing the combined SOC--TW bifurcation topology and the characteristic
spatiotemporal behavior of the concentration seen in numerical
simulations and experiments. The most important TW bifurcation features
are (i) a backwards Hopf bifurcation, (ii) a saddle node giving
rise to stable TWs, and (iii) the merging of the TW solution branch with the
SOC branch. Along the TW branch, (iv) both the phase velocity and the variance
of the concentration field decrease monotically in the same way.

{\it Model --- \/}
To derive such a model we started with an approximation for velocity
and temperature fields similar to that in the standard Lorenz model
\cite{L63}. We used however a more realistic improved version with a no
slip velocity field. To select and motivate an appropriate representation of
the concentration field we relied on a systematic analysis of the
concentration balance equations: the structure of the field components
occurring in a symmetry decomposition was investigated \cite{HL97b},
a separation into lateral mean fields and deviations thereof was used, and
the effect of the Soret coupling in the bulk of the fluid layer and at the
plates were quantitatively assessed for liquid mixture parameters
\cite{HL97a}. The concentration field truncation derived from these 
investigations consists of two linear and two nonlinear modes.
The resulting Galerkin model can be looked upon as minimal
for the description of convection in binary liquid mixtures since it
contains the Lorenz model, i.~e., the simplest truncation for the pure fluid,
and a minimal extension for mixtures. Previous extensions
\cite{C86a,AL87,LLMN88,L90} of the Lorenz model were too simple and therefore
failed in reproducing strongly nonlinear properties like, e.~g., the TW
saddle node. The present model is the first analytically manageable
approximation showing the above stated four characteristic properties (i)
-- (iv).

{\it Stationary convection --- \/}
Good agreement in the properties of
the SOC branch was found: stability threshold,
fluid parameters at the tricritical bifurcation, and position of the 
saddle node bifurcation. Furthermore, the approach of the SOC 
branch of a binary mixture to that of a pure fluid could be discussed
in the limit of large convective amplitudes.
Even the spatial variations of the
concentration field which is boundary layer dominated in SOCs are
reproduced in a way allowing good quantitative
agreement in the concentration variance. This holds
for both negative and positive $\psi$.

{\it Traveling waves --- \/}
The main results have to be seen in the description of TWs. Here, the model
shows that TWs occur only for negative Soret couplings in agreement with all
numerical simulations and experiments. It predicts
that the TW frequency is a direct measure for the "distance" of the
system from the pure fluid, i.~e., for the influence of the concentration
field. The "distance" can directly be read off the
bifurcation diagrams of flow intensity versus Rayleigh number
in two equivalent geometric
ways. This insight yielded a universal scaling relation between phase
velocity, convective velocity, and degree of mixing
of a TW which was confirmed in an impressive manner by different numerical
methods analysing the full field equations.
The derived scaling relation holds for all TW
states, stable or unstable independent of the bifurcation topology
\cite{HBL97}.

Linear convective properties are reproduced by
the model with high accuracy: oscillatory stability threshold, Hopf
frequency, and CT point. The same holds for the Rayleigh number at the
saddle node bifurcation. Our model shows also a SOC--TW transition 
to SOCs at the upper stationary stable branch and its
dependence on the separation ratio and Prandtl number is qualitatively
correct. However, its Lewis number dependence is unphysical since 
the concentration boundary layers which are responsible for the SOC--TW
transition \cite{BPS89} are represented in the model only in an
incomplete way. Otherwise, the spatial structure of the
concentration field in TWs 
--- frequency dependence of the concentration contrast between
the two plates and building up of plateaus (in the vertical direction) ---
is modelled in a quantitatively correct way by our
truncation.

The model allows to pinpoint the breakdown of an expansion of the TW
solution as a power series in the distance from the onset of
convection up to values where stable, strongly nonlinear TWs are
observed in experiments.
The related radius of convergence of the model is close to the numerically
determined one \cite{HBL97} marking the transition
from weakly to strongly nonlinear states \cite{HBL97}.
Hence, complex Ginzburg--Landau equations should not be expected to yield
reliable quantitative results for localized and extended TW states.

The results of the stationary and traveling states are brought together in
a phase diagram whose good agreement with numerical simulations
\cite{HL97a} can directly be inferred. Only in the SOC--TW transition
remarkable deviations are observed.

{\it Standing waves --- \/}
Our model gives insights into nonlinear
SW solutions in binary mixtures. It confirms earlier
weakly nonlinear results \cite{Kn86b,SZ93} like the initail slope.
In addition, it becomes possible to follow the SW branch,
which is everywhere unstable, up to regions with strongly nonlinear
oscillating amplitudes. A numerical determination of the unstable SW
solution of the full field equations is still lacking and an observation of
these states, say, by a control process is an experimental challenge.

\acknowledgements
This work was supported by the Deutsche Forschungsgemeinschaft. Fruitful
discussions with W.~Barten and P.~B\"uchel are gratefully
acknowledged.




\begin{references}
\bibitem{osc-inst} 
Contrary to statements found in the 
literature the oscillatory instability is not caused by concentration 
diffusion being small: even if the diffusive time scales for momentum, heat, 
and concentration are the same ($L = \sigma = 1$) there are oscillations.
They are caused by a 
sufficiently strong "off diagonal" Soret coupling between the degrees of 
freedom of concentration and temperature and their feedback into the 
buoyancy.
\end{references}
\end{document}